# A critical comparison of methods for the determination of the ageing sensitivity in biomedical grade yttria stabilized zirconia

Sylvain Deville [1], Laurent Gremillard [2], Jérôme Chevalier [1,#], Gilbert Fantozzi [1]

[1] Materials Science Department, National Institute of Applied Sciences (GEMPPM-INSA), Associate Research Unit 5510, 20 av. A. Einstein, 69621 Villeurbanne, France

[2] present address: Materials Science Division, LNBL, Berkeley CA 94720,USA

[#] corresponding author: e-mail: jerome.chevalier@insa-lyon.fr, Tel: +33 4 72 42 61 25, Fax: +33 4 72 43 85 28

**ABSTRACT**

Since the recent failure events of two particular series of zirconia femoral heads for total hip replacement prosthesis, a large decrease in the use of zirconia ceramics for orthopaedic implants has been observed. In spite of the biomedical success of this material during the last ten years, this was required for safety reasons, until the cause of the failures is known. It has been shown that these failures were related to the low temperature hydrothermal degradation (also known as ageing). Thus it is crucial to better understand the ageing behaviour, in order to be able to assess its importance and then control it if required. In this paper, various techniques relevant to assess the hydrothermal degradation sensitivity of biomedical grade yttria stabilized zirconia are discussed and compared. The expected outputs of conventional methods, i.e. X-Ray diffraction and scanning electron microscopy are examined. More recent methods like optical interferometry and atomic force microscopy are presented, with their respective benefits and drawbacks. An up to date comparison of these different techniques is provided, and their use for ensuring the long term reliability of a particular batch of zirconia in terms of ageing degradation is demonstrated.

## INTRODUCTION

The requirements of biomaterials for total hip replacement are very demanding [1]. To ensure the long term reliability and thus the clinical success of an implant, the selected materials must meet a unique combination of biocompatibility and mechanical properties. Total hip replacement has gone through half a century of researches and improvements, during which a wide variety of materials and materials combinations were examined. Until some recent failure events [2,3] of two particular series of zirconia femoral heads, one of the most performing solutions achieved was the use of yttria stabilized tetragonal zirconia (YTZP) for the femoral head [4,5]. Its excellent biocompatibility, high fracture toughness, high strength and low wear rates [6,7] made YTZP very attractive for this application.

The analyses of the broken femoral heads pointed to one of the drawbacks of yttria stabilized zirconia, its sensitivity to low temperature degradation, also referred to as *ageing,* which can have a deleterious effect on the service performance of YTZP components. This phenomenon has been investigated for the last thirty years, and is now fairly well established (when submitted to hydrothermal and / or mechanical stresses, the metastable tetragonal zirconia phase may transform to the stable monoclinic structure [8-11]). The recent failures have nonetheless alarmed the biomedical community. Though very concerning for the patients, they were limited to two particular batches. Their origin has been related to an *untimely, accelerated* and *unexpected* ageing of the heads, starting from the inner surfaces. This phenomenon was due to an accidental modification of the processing route. However, these events have had tremendous consequences on the use of zirconia as a bioceramic, since it has decreased dramatically. Many surgeons came back to alternative and sometimes less performing materials solutions. On the other hand, it is worth mentioning that most of the YTZP femoral heads have now been implanted for more than 15 years without any critical problem being reported. Though most of the factors affecting the ageing sensitivity were believed to be known, these events clearly proved there is room for new factors to be assessed. Knowing these new factors, the long term reliability of the prostheses can be ensured, provided that a careful analysis of their degradation sensitivity is carried out first.

The monoclinic phase is the stable structure of zirconia ceramics at room temperature. When stabilized with oxides such as yttria, ceria, magnesia or calcia, the zirconia ceramics can retain their high temperature tetragonal structure, which is metastable at room temperature. The transformation to the monoclinic phase may occur when an external mechanical stress is applied, and the resulting volume increase can slow down the crack propagation, leading to the very good fracture toughness and crack propagation resistance [12]. However, the transformation is also triggered by hydrothermal stresses [8,10] and can thus occur *in vivo* on the surface of implants, leading to their roughening and microcracking. This low temperature transformation in the presence of water is commonly called ageing. Ageing is schematically described in figure 1. The transformation starts first in isolated tetragonal grains on the surface. When a tetragonal grain transforms, the volume increase accompanying the phase transformation leads to stresses concentration in the surrounding zones and extensive microcracking. This stage corresponds to the formation of the so called *monoclinic spots* [13]. The stresses may trigger the transformation of neighbouring grains, while microcracks allow water penetrating into the material, so that the size of the monoclinic spot is increasing. Concurrently new monoclinic spots are formed elsewhere at the surface. The transformation is thus propagating from near to near, from the surface into the bulk, by a nucleation and growth process. Consequently the classical sterilization procedure performed in steam at 134°C did have a detrimental effect on long term stability and is now forbidden for zirconia [14]. Considering the ageing kinetics *in vivo*, the surface layer concerned by the transformation after several years of implantation should be limited to a few microns. However, it has been demonstrated that the transformation kinetic is very sensitive to a number of microstructural features. By modifying some of these parameters, the kinetics can be shifted by several orders of magnitude, as it has been the case in the two problematic series. Hence, assessing the ageing sensitivity of every batch of zirconia femoral heads is of prime importance.

The transformation has been proven to propagate by a nucleation and growth process starting at the surface and continuing on the surface and into the bulk. Therefore, characterizing the ageing behaviour and sensitivity necessitates accurate observation techniques of the transformed phase at the surface. Before starting new investigations to elucidate previously unaddressed factors, an up to date knowledge of the experimental methods that may be used to investigate ageing is required. Such a state of the art and comparison have never been made before. The objective of the present work is to conduct a critical comparison of new methods (optical interferometry (OI) and atomic force microscopy (AFM)) and well established methods (X-ray diffraction (XRD) and scanning electron microscopy (SEM)) for assessing the ageing sensitivity of YTZP and to describe the corresponding outputs. The advantages and drawbacks of each method are discussed.

## MATERIALS AND METHODS

*Materials processing*

The materials used in these experiments were processed from an atomised 3 mol.% $Y_2O_3$ zirconia powder (TZ3Y, Tosoh, Tokyo, Japan), to obtain biomedical grade materials, according to the ISO 13356:1997 normalisation. The microstructural characteristics are described in table I with comparison to ISO 13356. Bulk samples were polished for subsequent analysis using diamond based products, down to roughness values ($R_a$) below 3 nm.

A typical microstructure is shown in figure 2. Samples were thermally etched for 20 min at 1450°C in order to reveal the grain boundaries. The microstructure is homogenous without any important microstructural defects. An average grain size of 0.5 µm was measured by the linear intercept method (LIM) [15], as recommended by the ISO. It is worth noticing that due to stereological effects, the LIM grain size is smaller than the real grain size. A correction factor of $4/\pi$ is usually used to obtain the real grain size [16]. The next generation of ISO should maybe take into account these effects and specify the real grain size, indication that makes more sense, from a microstructural point of view. The density was measured by the Archimedes method.

*Experimental techniques*

**Low temperature autoclave ageing** The transformation being both thermally activated [13] and accelerated by the presence of water, samples were put in autoclave in steam during controlled times at 134°C, at a 2 bars pressure, in order to induce the phase transformation at the surface. This temperature was chosen for comparison with the standard steam sterilisation procedure [14].

**X-Rays diffraction** Since ageing is resulting from a phase transformation, the first possibility to quantitatively follow the transformation is to measure the phase fraction evolution by X-Ray diffraction (XRD). XRD data were collected with a θ-2θ diffractometer using the Cu-Kα radiation. Diffractograms were obtained from 27° to 33°, at a scan speed of 0.2°/min and a step size of 0.02°. The monoclinic phase fraction $X_m$ was calculated using the well known Garvie and Nicholson method [17]:

$$X_m = \frac{I_{m(\bar{1}11)} + I_{m(111)}}{I_{m(\bar{1}11)} + I_{m(111)} + I_{t(101)}} \quad (1)$$

where $I_t$ and $I_m$ represent the integrated intensity (area under the peaks) of the tetragonal (101) and monoclinic (111) and (-111) peaks. The monoclinic volume fraction is then given by:

$$V_m = \frac{1.311 X_m}{1 + 0.311 X_m} \quad (2)$$

**Scanning Electron Microscopy (SEM)** The propagation of the transformation into the bulk was followed by SEM (Philips XL20 with an accelerating voltage of 10 kV), on cross section samples [18]. A Philips XL20 microscope was used, with an accelerating voltage of 10 kV. The samples were coated with a 15 nm gold layer prior observation to make their surface conductive.

**Optical interferometry (OI)** OI (*Phase Shift Technology*) was used to investigate the surface degradation kinetics at a microscopic scale (lateral resolution around 2 µm, height resolution smaller than 1 nm). OI allows observing the surface relief induced by the apparition and growth of monoclinic spots. No specific sample preparation is needed. Polished samples or explanted femoral heads surfaces may be observed straightaway.

**Atomic force microscopy (AFM)** AFM is one of the newest microscopy techniques. As it has been introduced only a few years ago in materials science, very few observations of ceramic surfaces using AFM have been reported in the literature so far. Recently reported results [19,20] have nonetheless drawn the attention on the potentialities offered by AFM to investigate surface modification of zirconia containing ceramics. The main interest of AFM relies in its ability to provide informations of the surface state at the nanometer scale. The vertical resolution can be as low as atomic scale. Considering the scale at which the transformation occurs, AFM appears to be an extremely powerful tool. Thus, A DI3100 microscope from *Digital Instruments Inc.* was used in contact mode, using oxide sharpened silicon nitride probes with an average scanning speed of 10 µm.s$^{-1}$, without any surface preparation.

**RESULTS AND DISCUSSION**

*Outputs of the experimental techniques*

XRD is the most commonly used method for the quantitative evaluation of the transformation kinetics. The figure 3 shows the time evolution of the monoclinic phase fraction as measured by XRD. The transformation rate increases progressively up to a maximum before decreasing, when the monoclinic fraction reaches a plateau value. Although the monoclinic fraction is smaller than 100%, the transformed fraction does not exhibit any further increase after about 12 hrs at 134°C. The sigmoïdal shape of the curve is characteristic of a nucleation and growth mechanism, as mentioned in the introduction. Using the Mehl-Avrami-Johnson (MAJ) formalism [21], the transformed monoclinic fraction *f* as a function of the ageing time *t* can be written:

$$f = 1 - \exp\left(-(bt)^n\right) \quad (3)$$

where n is related to the nucleation and growth conditions (specific of for each material) and b is a thermally activated term described by:

$$b = b_0 \exp\left(-\frac{Q}{RT}\right) \quad (4)$$

where $b_0$ is a factor specific to every nuance of zirconia, Q the activation energy of the transformation (about 106 kJ/mol, from Ref 13.), R the gas constant and T the temperature. Since the transformation is thermally activated, autoclave ageing tests can be performed to predict the long term behaviour at body temperature. It has been calculated that one hour of autoclave treatment at 134°C has theoretically the same effect as three years in vivo at 37°C [13]. A lot of heavy experimental tests of several years on animals or patients can therefore be avoided, although performing some of them remains undoubtedly necessary. The steam sterilisation procedure had therefore a very deleterious effect on the ageing behaviour of zirconia femoral heads, 2 hours of sterilization having roughly the same effect as 6 years in vivo, and was consequently forbidden a few years ago [14].

SEM can be used to investigate further propagation into the material. Since the transformation is accompanied by extensive microcracking, the transformed zones are easily taken away when polishing the surface, leaving holes where the material has transformed to its monoclinic structure. Therefore the observation of transformed zones under the surface is possible on cross section samples (fig. 4). The first micrograph was taken after 3 hrs of ageing at 134°C, and shows a transformed monoclinic spot that was intercepted by the cross section plane. At this time only this small zone was transformed, the remaining of the surface and the volume still being constituted of healthy, untransformed zirconia. When the ageing treatment time increases, the transformation propagates further into the volume. The micrographs show that the penetration depth of the transformation is about 4 µm after 11 hours and 11 µm after 16 hrs at 134°C in the observed material.

OI allows observing the relief change with a sub-nanometer vertical resolution. The volume increase accompanying the formation of the monoclinic phase can be clearly observed, (figure 5). At each point of the image, the brightness is proportional to the height above the average level of the surface. Since the phase transformation is accompanied by a volume increase, the monoclinic spots appear brighter. The acquired image contains a 3D description of the surface relief with a very precise vertical resolution. By carrying out repeated ageing experiments and observations, it is possible following the nucleation and growth of monoclinic phase as a function of the ageing treatment time. It is also possible quantifying the fraction of transformed surface by image analysis.

AFM is the newest technique applied to the observation of surface transformation of zirconia. If the vertical resolution is very similar to that of OI, its lateral resolution is much improved. The observation of monoclinic spots with an apparent diameter smaller than 2 µm is possible, which was not the case with OI. AFM images of the very first stages of the transformation are shown in Fig. 6. Like OI, AFM pictures provide a relief contrast, and the transformed zones appear *brighter* on the images. On a given sample, the same area can be observed after each autoclave treatment, enabling the determination of the evolution of the transformation. The transformation propagation can be followed at a nanometer scale (even partially transformed grains can be observed, see Figure 6). In addition, evolution of the relief induced by the transformation can be measured very accurately.

*Critical comparison of the experimental techniques*

The relevance domain of each technique is schematically described in Fig. 7 and 8, in regards of the results presented here. While XRD was traditionally used to follow quantitatively the transformation propagation, the limits of this technique will be exposed here. On one hand, no precise information can be obtained during the first stages of the transformation: the precision of the measurements is limited by the signal – noise ratio, especially at low transformed fraction. Also, a variability of the results can appear when analyses of different locations on the same sample are conducted. For these reasons, it is not possible obtaining reliable information by XRD for transformed fractions smaller than 5 %, i.e. for the first stages of the transformation. In addition, the XRD signal comes from the surface layer only, typically no more than the top few microns. The information provided by XRD is thus related to the near surface of the sample and not to the bulk. Moreover, as the X-ray probe is a few mm wide, no local information can be obtained. Therefore the XRD measurements characterise the overall sample behaviour. The limited penetration leads to the apparent saturation of the transformation when measured by XRD. Indeed, when the transformed layer is deeper than the X-ray penetration depth, the ageing can not be followed by XRD anymore. Finally, it should be mentioned that XRD is a non destructive method. This technique can be considered as the first step for investigating the ageing sensitivity of any particular batch of zirconia.

SEM observations of cross sections can be used to follow the transformation propagation into the volume. The ageing kinetics measured by XRD exhibited stagnation after 12 hours of treatment, while it is clear from the SEM observations that the transformation is still propagating into the volume of the material. Therefore this apparent stagnation is clearly related to the limited depth analysed by XRD. Cross sectional SEM observations avoid this drawback and enable following the transformation propagation as far as necessary below the surface. However, the fact that this technique is destructive constitutes a major disadvantage.

XRD and SEM were the two methods traditionally used to follow quantitatively the transformation. However, it is quite clear from the results presented here that they exhibit a limited resolution, in particular during the first stages of the transformation. In addition, the behaviour of the material during the first stages will determine the whole transformation kinetics. From a long term point of view, obtaining a precise knowledge of these first stages is necessary. Techniques providing an improved spatial resolution should therefore be considered.

In regards of these objectives, more recent techniques like optical interferometry (OI) and atomic force microscopy (AFM) are of great interest. The improved spatial resolution allows observing surface modifications during the first stages of the transformation, features that were not accessible with conventional techniques (SEM and XRD). One of the great advantages of OI is the non-destructive character of the technique, as compared to SEM. However, it does only provide information about the surface transformation. No information about the propagation within the volume can be obtained.

If precise predictions of the long term degradation kinetics are required, a variety of techniques must be considered. For instance, the numerical simulation of ageing kinetics [22] has been reported recently as an alternative method for the prediction of long term behaviour. Precise values of the nucleation and growth parameters are required by such modelling. Both OI and AFM can be fruitfully used to provide these parameters, with similar vertical resolutions. While optical interferometry allows observing large zones of the surface (e.g. useful to measure the nucleation rate), the technique suffers from a poorer lateral resolution. Monoclinic spots with an apparent diameter at surface smaller than 2 µm cannot be observed, while even partially transformed grains can be observed by AFM [20] [Figure 6]. On the other hand, large zones (120*160 µm) can be observed by OI, providing statistically reliable measurements. Hence, with its nanometer scale lateral and vertical resolutions, AFM is an alternative solution for measuring the transformation parameters in the very first stages of the transformation. This technique is however limited by the size of the zones to be scanned (typically no more than 100 µm x 100 µm). Additional experiments must be carried out to get statistically significant parameters. Both AFM and optical interferometry do not require specific sample preparation, as opposed to SEM, and are non-destructive techniques.

## CONCLUSIONS

The different experimental techniques that may be used for the characterization of the low temperature ageing phenomenon have been described and compared. The use of accelerated ageing tests in autoclave provides very valuable insights on the long term ageing behaviour of zirconia ceramics, in particular to compare different material solutions. SEM and AFM provide local observations of the degradation propagation, respectively in volume and on surface. XRD can be used for quantifying the degradation kinetics. If a numerical prediction of the degradation kinetics is considered, optical interferometry can be used with AFM to provide the numerical simulation parameters, in particular during the first stages of the transformation. The complete knowledge of the ageing kinetics of zirconia ceramics necessitates the use of all these techniques together, as none of them can describe the whole transformation process.

## ACKNOWLEDGEMENTS

Financial support of the Rhône-Alpes region and the European Union (GROWTH2000, project BIOKER, reference GRD2-2000- 25039). The authors would also like to thank the CLAMS for using the nanoscope, and Laure Notin and Stéphanie Michaud for their contribution to this work.

Figure 1: Scheme of the ageing process seen in cross section, showing the near to near transformation propagation, accompanied by extensive microcracking (dashed areas), creating new paths for the water to penetrate into the material. The volume increase and important shear induced by the transformation leads to a modification of surface relief.

Figure 2: Typical microstructure observed by SEM.

Figure 3: Transformation kinetics at 134°C measured by XRD, for several samples of the same material. Scattering of the results is obvious.

Figure 4: Cross sections of an aged sample, showing the initiation (monoclinic spot, after 3 hrs) and propagation of the transformation (after 11 hrs and 16 hrs). The transformed zone, with a high density of microcracks, is very brittle and easily taken away during the polishing stage, leaving a hole at surface, as seen here in a cross section.

Figure 5: Observation by optical interferometry of surface transformation of the same area after 3 hrs and 7 hrs at 134°C. Nucleation and growth of monoclinic spots is clearly observed. The surface relief induced by the monoclinic spots growth can be individually followed when the apparent spot size at surface reaches a few microns. Micrographs extracted from Ref [13]. Reprinted with permission of the American Ceramic Society, www.ceramics.org Copyright 1999. All rights reserved.

Fig. 6: Atomic force microscopy observations of surface transformation of the same area, after 8 min and 32 min at 134°C. Even partial transformation of the grains can be observed. The long dark straight lines are polishing residual scratches.

Fig. 7: Relevance domains of the experimental techniques as a function of the transformation stage, and penetration depth of each technique.

Fig. 8: Relevance domains of the experimental techniques as a function of the localisation of the transformation and penetration depth.

Table I: Comparison of the materials properties of this study to the ISO requirements.

| Property | ISO 13356:1997 | Materials of this study [#] |
|---|---|---|
| $ZrO_2$ + $HfO_2$ | > 94.05 % | > 94.7 % |
| $HfO_2$ | ≤ 0.5 % | ≤ 0.5 % |
| $Y_2O_3$ | 4.95 ± 0.45 % | 5.12 % |
| $Al_2O_3$ | < 0.5 % | 0.1 % |
| Other impurities | < 0.5 % | < 0.02 % |
| Density | 6.00 g.cm$^{-3}$ | 6.05 g.cm$^{-3}$ |
| Linear intercept grain size | 0.6 µm | 0.5 µm |
| Surface roughness (Ra) | 20 nm | < 3 nm |

[#] chemical composition provided by the manufacturer

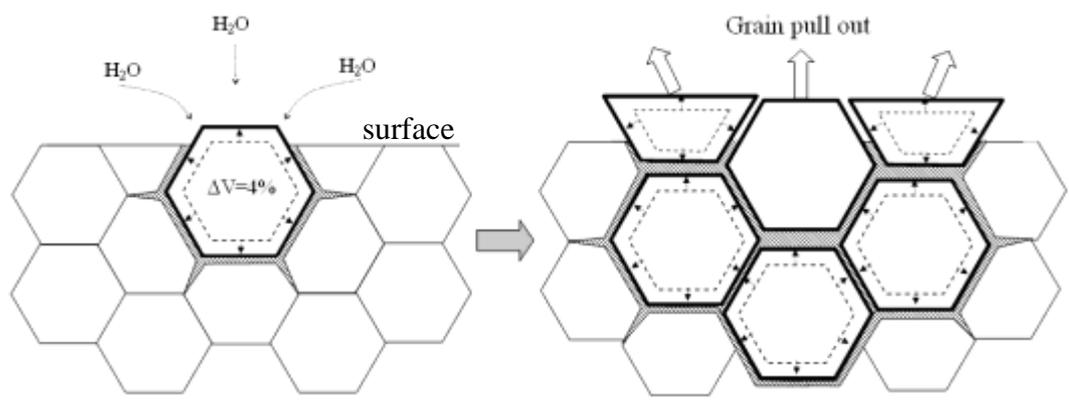

Fig. 1

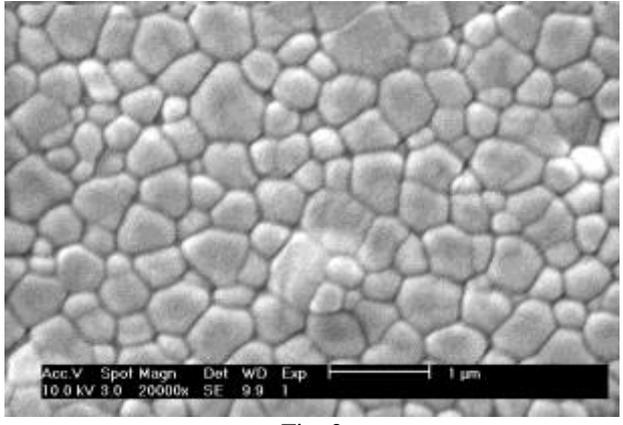

Fig. 2

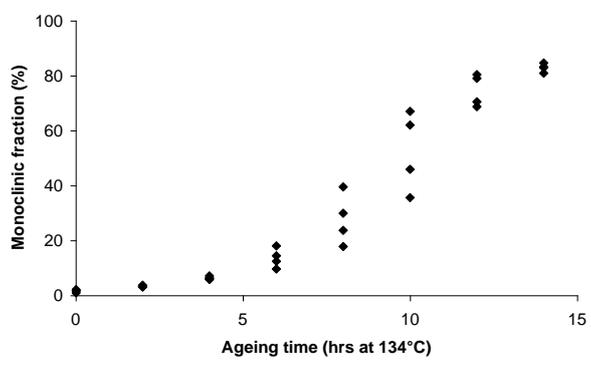

Fig. 3

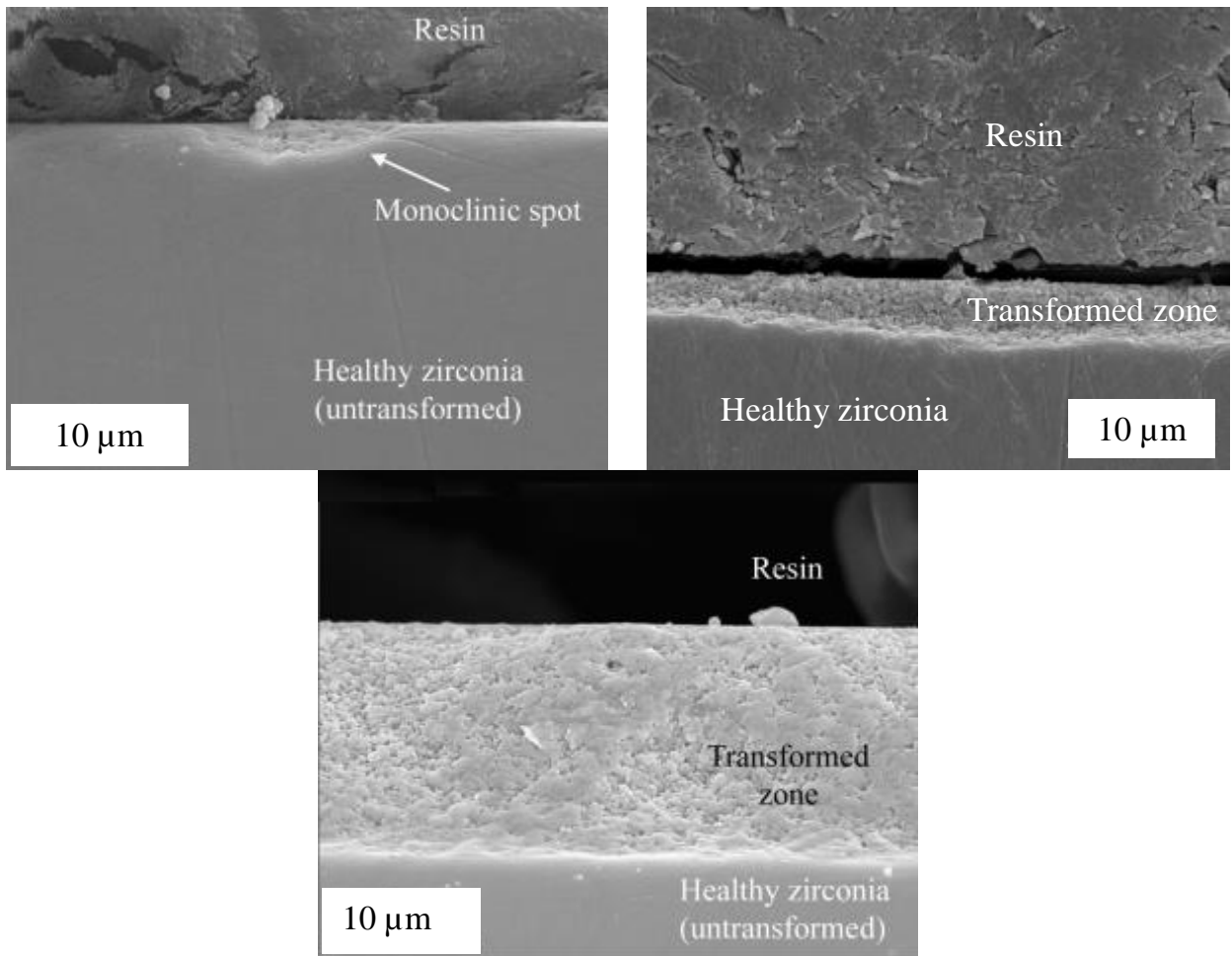

Fig. 4

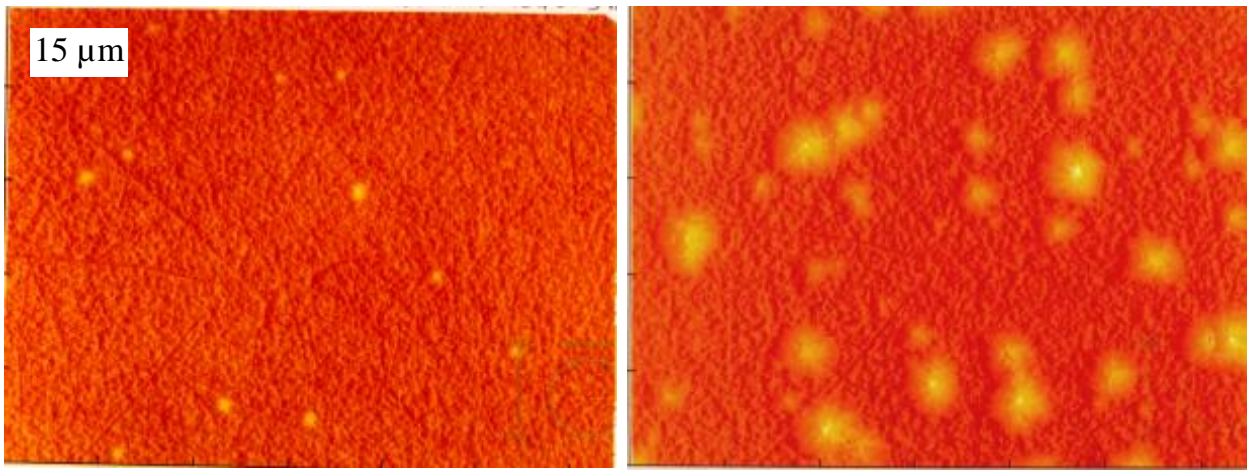

Fig. 5

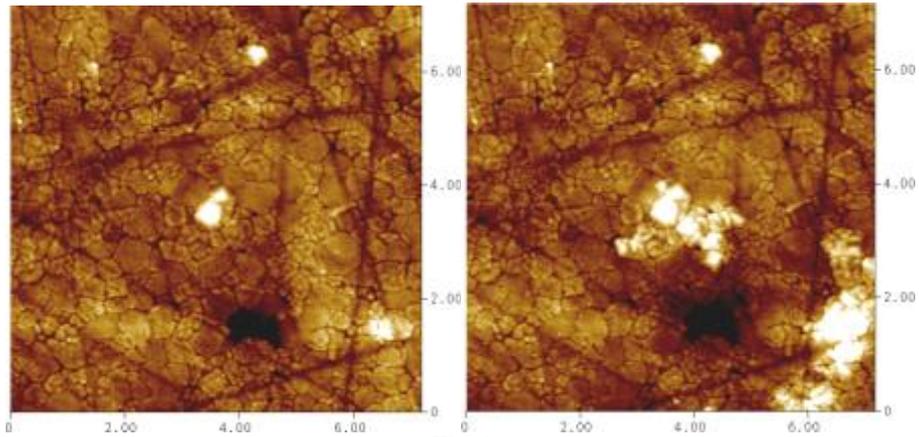

Fig 6

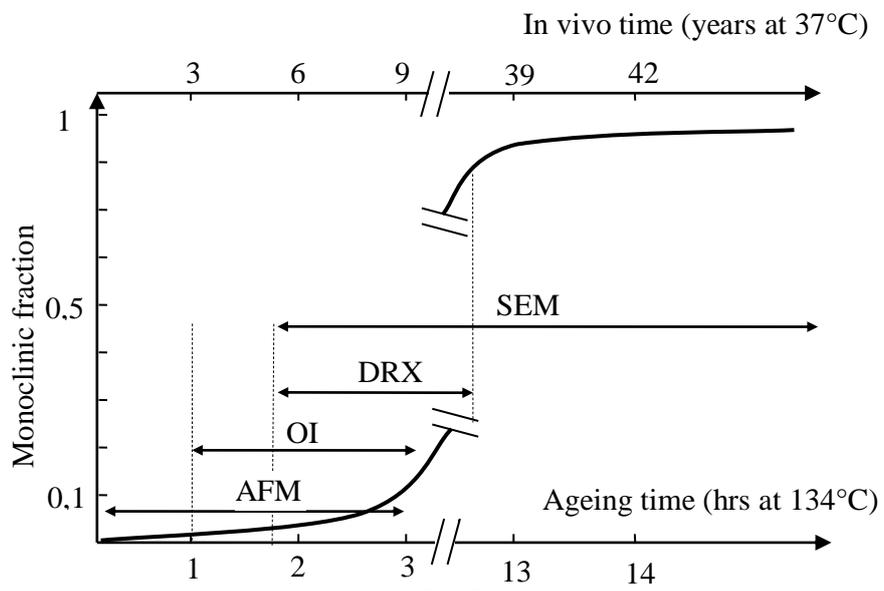

Fig. 7

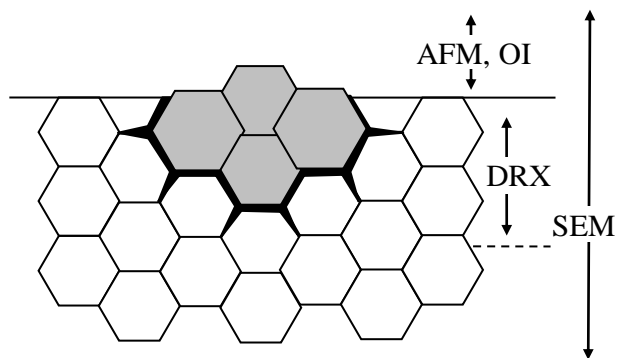

Fig. 8